\documentclass[]{spie}  

\newcommand\arcsec{\mbox{$^{\prime\prime}$}}%

\let\OLDitemize\itemize
\renewcommand\itemize{\OLDitemize\addtolength{\itemsep}{-5pt}}

\usepackage{amsmath,amsfonts,amssymb}
\usepackage{graphicx}
\usepackage[colorlinks=true, allcolors=blue]{hyperref}

\title{Optimal scheduling and science delivery of spectra for millions of targets in thousands of fields: the operational concept of the Maunakea spectroscopic explorer (MSE)}

\author[a]{Nicolas Flagey}
\author[b]{Alan McConnachie}
\author[a]{Kei Szeto}
\author[c]{Patrick Hall}
\author[a]{Alexis Hill}
\author[a]{Calum Hervieu}

\affil[a]{Canada-France-Hawaii Telescope Corporation, 65-1238 Mamalahoa Hwy, Kamuela HI 96743, USA}
\affil[b]{NRC Herzberg, Dominion Astrophysical Observatory, 5071 West Saanich Road, Victoria, British Columbia, Canada}
\affil[c]{Department of Physics and Astronomy, York University, Toronto, ON M3J 1P3, Canada}

\authorinfo{Further author information: (Send correspondence to N.F.)\\N.F.: E-mail: flagey@cfht.hawaii.edu}

\begin{document} 
\maketitle

\begin{abstract}
The Maunakea Spectroscopic Explorer (MSE) will each year obtain millions of spectra in the optical to near-infrared, at low ($R\simeq 3000$) to high ($R\simeq 40000$) spectral resolution by observing $>$3000 spectra per pointing via a highly multiplexed fiber-fed system. Key science programs for MSE include black hole reverberation mapping, stellar population analysis of faint galaxies at high redshift, and sub-km/s velocity accuracy for stellar astrophysics.

The architecture of MSE is an assembly of subsystems designed to meet the science requirements and describes {\it what} MSE will look like. In this paper we focus on the operations concept of MSE, which describes {\it how} to operate a fiber fed, highly multiplexed, dedicated observatory given its architecture and the science requirements.

The operations concept details the phases of operations, from selecting proposals within the science community to distributing back millions of spectra to this community. For each phase, the operations concept describes the tools required to support the science community in their analyses and the operations staff in their work. It also highlights the specific needs related to the complexity of MSE with millions of targets to observe, thousands of fibers to position, and different spectral resolution to use. Finally, the operations concept shows how the science requirements on calibration and observing efficiency can be met.
\end{abstract}


\section{INTRODUCTION}
\label{sec:intro}  

The Maunakea Spectroscopic Explorer (MSE) is a project to upgrade the 3.6-meter telescope and instrumentation of the Canada-France-Hawaii Telescope (CFHT) into a 11.25-meter telescope equipped with fiber-fed spectrographs dedicated to optical and near-infrared (NIR) spectroscopic surveys. The current baseline for MSE is that of a prime focus, 10-meter effective aperture telescope feeding a bank of low and moderate spectral resolution spectrographs (LR, R$\sim$3000 and MR, R$\sim$6000) located on platforms, as well as high spectral resolution spectrographs (HR, R$\sim$20000 and R$\sim$40000) located in the more stable pier of the telescope. The 1.5 square degree field of view of MSE will be populated with 3249 fibers of 1\arcsec\ diameter allocated to the LMR spectrographs, and 1083 fibers of 0.8\arcsec\ diameter for the HR spectrographs. The fibers will all be positioned with tilting spines, with both LMR and HR fibers being available at all time.

At the previous SPIE Astronomical Telescopes and Instrumentation meeting, the status and progress of the project were detailed in Ref.~\citenum{Murowinski2016} while an overview of the project design was given in Ref.~\citenum{Szeto2016} and the science based requirements were explained in Ref.~\citenum{McConnachie2016}. An update of the project at the end of conceptual design phase is presented this year in Ref.~\citenum{Szeto2018a} with a review of the instrumentation suite in Ref.~\citenum{Szeto2018b}. Other papers related to MSE are focusing on: the summit facility upgrade (Ref.~\citenum{Bauman2016, Bauman2018}), the telescope optical design for MSE (Ref.~\citenum{Saunders2016}), the telescope structure design (Ref.~\citenum{Murga2018}), the design for the high-resolution (Ref.~\citenum{Zhang2016, Zhang2018}) and the low/moderate-resolution spectrograph (Ref.~\citenum{Caillier2018}), the top end assembly (Ref.~\citenum{Mignot2018, Hill2018b}), the fiber bundle system (Ref.~\citenum{Venn2018, Erickson2018}), the Sphinx fiber positioner system (Ref.~\citenum{Smedley2018}), the systems budgets architecture and development (Ref.~\citenum{Mignot2016, Hill2018}), the observatory software (Ref.~\citenum{Vermeulen2016}), the spectral calibration (Ref.~\citenum{Flagey2016a, McConnachie2018a}), the throughput optimization (Ref.~\citenum{Flagey2016b, McConnachie2018b}), the injection efficiency (Ref.~\citenum{Flagey2018ie}), and the observing efficiency (Ref.~\citenum{Flagey2018oe}).

MSE will be a facility dedicated to spectroscopic surveys. Multiple surveys will be executed in parallel to make the best use of the observatory capabilities, with both LMR and HR spectrographs simultaneously available at all time and sharing the telescope's field of view. While the surveys are envisioned to be multi-year large programs (LP), we expect that some observing time might be allocated to small programs (SP) that will likely be more focused, require smaller amounts of telescope time or number of fibers, and typically lead to more rapid publications. Even though MSE will be a significantly different facility than CFHT, the operations of MSE will benefit from CFHT's experience in queue service observations and from the transition that has already happened at CFHT with an increased importance of LPs and survey oriented operations.

In this paper we focus on the operations concept of MSE: how MSE will be operated by its support staff to provide all that the science community needs to submit proposals, define targets, and retrieve data that have met the science requirements in terms of quality and quantity. As such, we start by presenting the organizational structure of MSE in section \ref{sec:org}. Then, we detail the operations concept various phases of operations (see section \ref{sec:phases}). Section \ref{sec:sched} discusses the complexity of the scheduling for a facility like MSE where multiple survey programs comprising millions of targets and using different spectral resolution will coexist within a field of view populated by thousands of fiber positioners. We then describe the tools needed at each step to support the operations staff and the science community (see section \ref{sec:tools}). The science requirements on observing efficiency and calibration and their flow down to the operations concept are then briefly presented in sections \ref{sec:obseff} and \ref{sec:calib}.

\section{Organization}
\label{sec:org}

The MSE staff is envisioned as being divided into five functional groups: Administration \& Finance, Facility Operations, Instrument Operations, Software Operations, and Science Operations, each defined by their skill sets, specialties, and role in the organization. The groups share observatory resources and work collaboratively in support of each other. This section summarizes their role.

\subsection{ADMINISTRATION \& FINANCE GROUP}
The Administration \& Finance Group (AFG) is responsible for all finance, human resources, and administrative support and oversight activities. By managing these support functions, the AFG enables employees and managers throughout MSE to focus completely on the operational and scientific needs of the organization without administrative distractions.

\subsection{FACILITY OPERATIONS GROUP}
The Facility Operations Group (FOG) is responsible for the physical plants and equipment, including mechanical plants (heating, ventilation, air conditioning, coolant, chiller, compressed air and hydraulics), electrical plant, telescope/enclosure drives, and other industrial equipments and safety hardware. The FOG maintains the performance, reliability and availability of these systems to enable science operations on a daily basis. In addition, the FOG supports instrument operations with their “industrial” skills, such as staging, rigging and lifting with the crane systems, and servicing and maintaining the telescope optics, instrument subsystems on the telescope, and the spectrographs.

\subsection{INSTRUMENT OPERATIONS GROUP}
The Instrument Operations Group (IOG) is responsible for the telescope optics and telescope mounted subsystems, both physical hardware and control electronics, including but not limited to primary mirror (M1), fiber positioner and metrology system (PosS and FPMS), and spectrographs (LMR and HR). The IOG maintains the performance, reliability and availability of these systems to enable efficient science operations on a nightly basis. In addition, the IOG operates specialized laboratories to service, repair and test observatory components in the areas of optics, fiber optics, mechanics, opto-mechanics, electronics and detectors, etc.

\subsection{SCIENCE OPERATIONS GROUP}
The Science Operations Group (ScOG) is responsible for all science-related aspects of the acquisition, processing and distribution of MSE science data. The ScOG will comprise astronomers hired by MSE to work in the Waimea headquarters to provide a scientific support to instrument, facility, and software operations. They will maintain an active research portfolio that includes the use of MSE science data. They will be expected to be a major interface between the Observatory and its user community.

\subsection{SOFTWARE OPERATIONS GROUP}
The Software Operations Group (SOG) is responsible for the MSE  data by maintaining and operating the observatory computer network, hardware and software. The SOG supports a computer network architecture that performs two major system functions: observatory and program execution. The Program Execution System Architecture (PESA) is optimized for interfacing with scientists to facilitate pre-observation planning and post-observation pipeline for data analysis, processing, storage and distribution. The Observatory Execution System Architecture (OESA) is optimized for enabling nighttime remote observations, coordinating and controlling the MSE subsystems to perform observations in an efficient and safe manner. The OESA also includes control of engineering systems that are used by technical staff to interact with the observatory systems. In addition, the SOG maintains an observatory level safety system, which is an independent hardware framework for the protection of personnel and equipment, and the SOG also supports a parallel status server, which collects and monitors components' “health” information for safety and maintenance purposes.

\section{Phases of operation}
\label{sec:phases}

\begin{figure}
\caption{\label{fig:phases} Summary of the different phases of operations for MSE with the tools needed for each phase.}
\includegraphics[width=\linewidth]{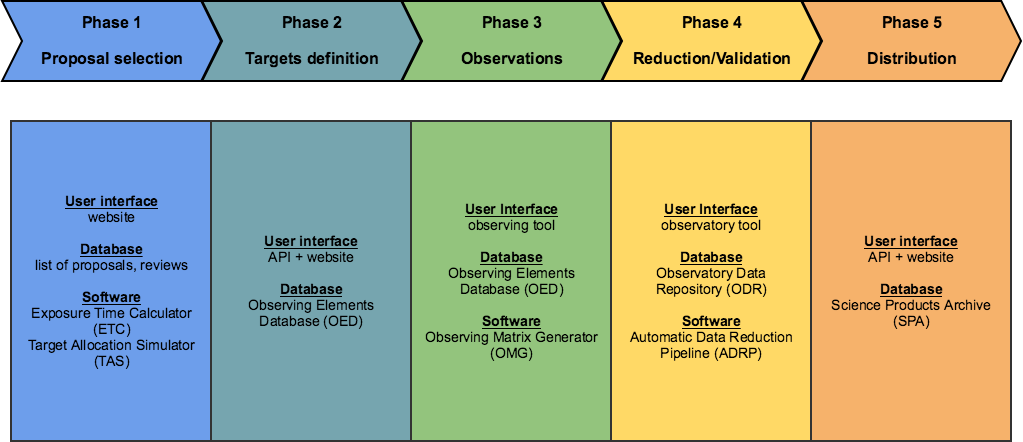}
\end{figure}

The operations at MSE will follow the usual phases for a ground based astronomical facility: observing programs will be selected (Phase 1 or PH1); approved programs will supply targets and instrument configurations will be specified (Phase 2 or PH2); observations will be executed (Phase 3 or PH3); data will be reduced and quality analysis performed (Phase 4 or PH4); data products will be distributed and archived (Phase 5 or PH5). Figure \ref{fig:phases} shows the different phases and introduces the Tools (i.e. software, databases, and user interfaces) required for each phase.

\subsection{Phase 1}
\label{sec:ph1}

The Phase 1 (PH1) refers to the step -- and associated tools -- during which investigators will request the use of the telescope to observe large samples of targets, based on a scientific justification and a technical justification. PH1 will process survey programs that will surely encompass thousands to millions of targets. Hereafter we describe the PH1 of MSE in broad terms as the details of the selection process will be defined at a later stage by the partnership.

\begin{figure}[h]
\centering
\caption{\label{fig:phase1} Activity diagram for PH1 summarizing the activities related to proposals submission and selection.}
\includegraphics[width=.75\linewidth]{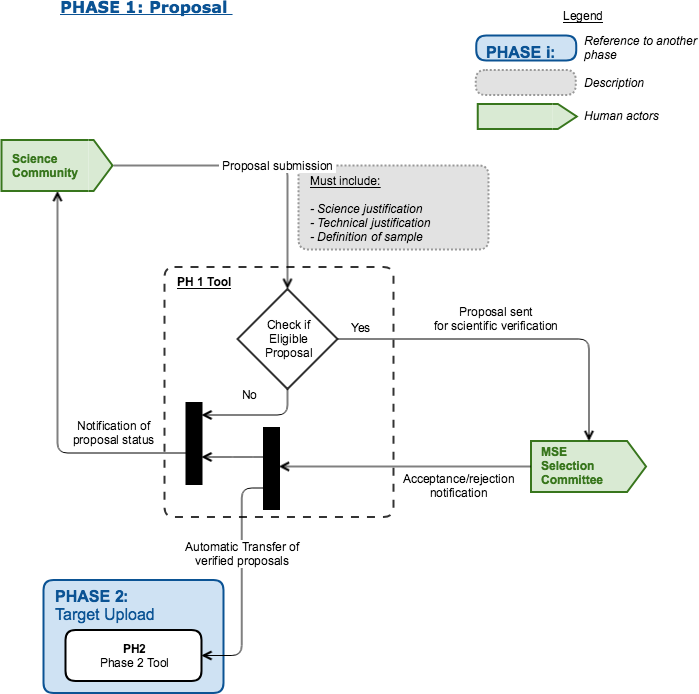}
\end{figure}

We expect that calls for proposals will be issued, and observing programs selected, at regular intervals. Depending on the type of programs (SP or LP) the selection could happen every few months or every few years. The PH1 Tool will be used to receive these proposals, which will contain a science justification, a technical justification, and a simple definition of the sample to observe. The PH1 Tool will automate and simplify the reviewing process of the proposals by a selection committee. Finally, it will automatically transfer to the Phase 2 the programs that have been approved for observations. Figure \ref{fig:phase1} shows an activity diagram for PH1.

To facilitate the scheduling of the observations and avoid duplicates, we envision that observing programs will define their sample of targets in a clear but simple way in PH1. Given the expected number of targets an observing program will surely request to observe (thousands to millions), we do not envision that a complete list will be provided in PH1. Instead, there will be a way for each submitted program to broadly define their targets sample with a list of criteria (e.g. boundaries of a region in the sky, magnitude range, object type). These criteria will later be used when targets are added in PH2 (see section \ref{sec:ph2}) to automatically verify that each program is observing targets they were approved for. The rules (flexibility, accuracy) about the criteria provided by the observing programs during their PH1 will be defined at a later stage in the project. The composition of the selection committee and the exact rules for the proposals submission and selection will also be defined at a later stage although it will surely comprise members of the Science Team representing the partnership and members of the ScOG.

A couple of other tools will be very useful for PH1: the Exposure Time Calculator (ETC\footnote{\url{http://etc-dev.cfht.hawaii.edu/mse/index.html}}) and the Target Allocation Simulator (TAS\footnote{\url{http://etc-dev.cfht.hawaii.edu/mse/alloc.html}}). The ETC simulates the observatory performance in terms of sensitivity and predicts the time required to reach a given SNR for a given target (point source or extended source, magnitude, template) and for a given set of weather conditions (seeing, sky brightness). The TAS simulates the observatory in terms of fiber to target allocation and predicts how many targets can be allocated to fibers in a given field for a given spectrograph. This will help the science team define their sample based on the estimated amount of time required to observe their list of targets. Sections \ref{sec:etc} and \ref{sec:tas} describe the ETC and the TAS in more details, respectively.

\subsection{Phase 2}
\label{sec:ph2}

\begin{figure}[h]
\centering
\caption{\label{fig:phase2} Activity diagram for PH2 summarizing the activities related to target definition.}
\includegraphics[width=.8\linewidth]{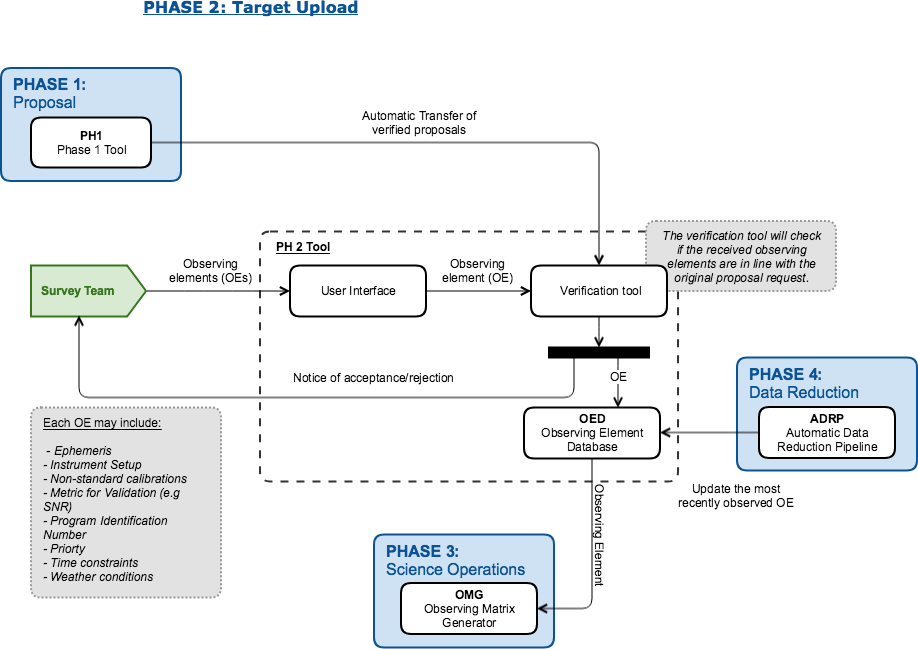}
\end{figure}

Phase 2 (PH2) corresponds to the step when the teams leading the approved observing programs (survey teams) will give all the required information about the observations they wish to obtain with MSE. Survey teams will provide Observing Elements (OE), using the PH2 Tools, which will populate the Observing Elements Database (OED). Figure \ref{fig:phase2} shows an activity diagram for PH2.

\subsubsection{Observing Elements}
The OEs will be defined by:
\begin{itemize}
\item A unique set of {\bf coordinates}: the coordinates should be precise enough to maximize the injection efficiency at the fiber input and will be specified in the Gaia reference coordinate system. All catalogs used by MSE will use that same reference frame.

\item {\bf Ephemeris} for proper motion objects: the ephemeris should follow the same convention and have the same precision as that required for the coordinates. They should be provided for a period of time long enough to cover the duration of the program. Fast moving targets will not be suitable for MSE, although open-loop positioning of the fibers could enable observations of targets with apparent motion no greater than about 0.5\arcsec\ per hour.

\item A {\bf spectrograph} (LR, MR, or HR): each OE will correspond to only one instrument setup. Some targets might require the use of multiple spectrographs (e.g. follow-up in HR mode of a previous observations in MR mode). The survey team will thus create several OEs for that given target, each with their own instrument setups.

\item Any {\bf non-standard calibrations} required for that observation. There will be a set of default calibrations established for each of the LR, MR, and HR spectrographs (see section \ref{sec:calib}). Any calibration not covered by these defaults will be specified in PH2 and will have been justified in PH1.

\item A metric for {\bf validation}: the observing programs will provide a metric for validation of the OE. For instance, a SNR goal, along with a magnitude, at a given wavelength, will be defined for each target based on calculations done by the Exposure Time Calculator (ETC). The observed SNR will be measured by the reduction pipeline for each target at the wavelength given by the program. There could be programs, however, for which the goal will be to obtain a SNR goal averaged over all the targets in a given field or over the entire sample.

\item A {\bf status flag} with initial value of "unobserved" that will switch to "observed" and "completed" as observations are made for that OE (see section \ref{sec:adrp})
.
\item A program {\bf identification number} to link each OE to a survey.

\item A {\bf priority}, internal to the observing program, between 0 (lowest) and 9 (highest). It will help schedule the OE. Internal priorities will be combined with global priorities set by MSE when allocating targets from multiple programs in the same field.

\item Optionally, a set of {\bf Time constraints} (e.g. start, finish, repeats), a set of {\bf Weather constraints} (e.g. sky background, seeing), a set of {\bf Setting constraints} (e.g. observe at the same time as another target). For example, some targets might need to be observed repeatedly, others might need to be observed only within some preferred time windows, and yet others might need to be observed in sequence with a delay in between. Because MSE will be a dedicated survey facility, weather constraints will need to be justified and approved in PH1. Additional constraints may be allowed on the settings of the observations. For instance, investigators may want to always observe a given target with the same fiber, or they may want to observe several targets at the same time. Those constraints will need to be justified in PH1 and approved by MSE.

\end{itemize}

\subsubsection{PH2 Tool and OED}

Observing programs will generally provide large numbers of targets (thousands to millions). An automatic process will be developed to simplify the definition of the OE and verify that each OE is valid. The PH2 Tool will allow the creation of catalogs of OEs, e.g. via an Application Programming Interface (API). There will be some tests to determine if an OE is acceptable (e.g. targets will likely need to be fainter than a given level), rejected targets will be flagged, and the investigators will be notified. There will surely be targets of opportunity (ToO), or other targets defined on a nightly basis after observations performed at other observatories (e.g. LSST). The investigators will therefore need a way to add targets at any time. Modifications to an OE will also be possible, as long as the OE has never been queued for observation (see section \ref{sec:omg}).

The Observing Element Database (OED) will serve as the repository for all OEs. Access to the OED will be limited to the authorized users: the survey teams will have access to their OEs, while the MSE operations staff will likely have access to all the OED. A common user interface will be developed for all OED access.

\subsection{Phase 3}
\label{sec:ph3}

In Phase 3 (PH3), the information provided in PH2 is translated into commands for the observatory, and the observations are executed. PH3 is a complex phase. The Observing Matrix Generator (OMG) will use the information in the OED and the current conditions, monitored in realtime via sensors, to generate the schedule of observations. All the information about the next observation will be sent to all the subsystems to configure them and move them as necessary. More details about Scheduling are given in section \ref{sec:sched}.

\begin{figure}
\caption{\label{fig:phase34} Activity diagram for PH3 and PH4 summarizing the activities related to observations execution, data reduction, and data validation.}
\includegraphics[width=\linewidth]{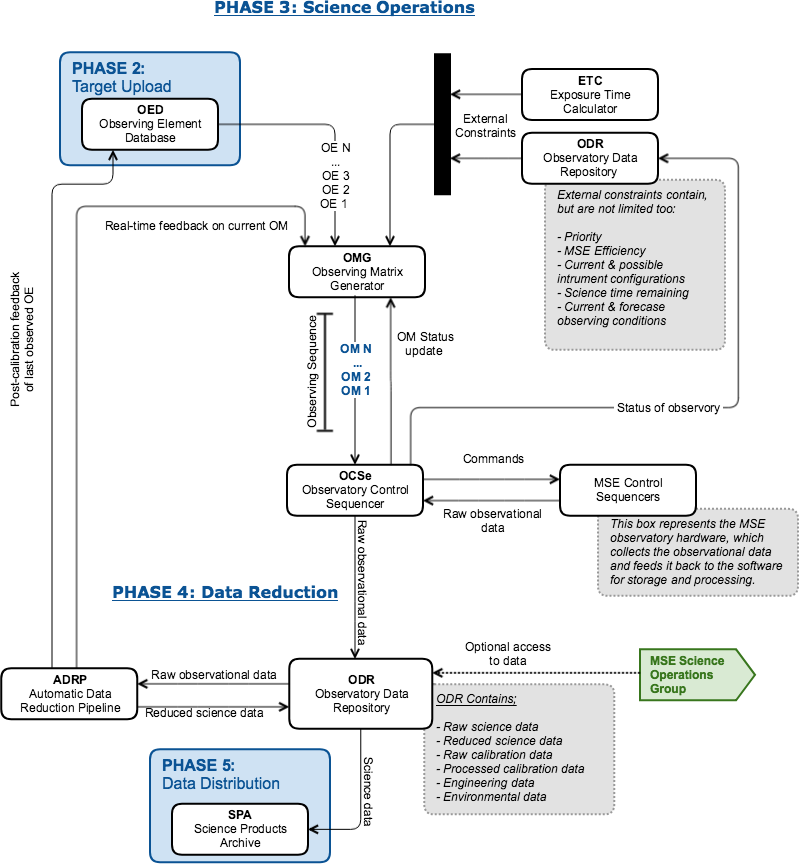}
\end{figure}

While the observations are executed, the observing conditions (e.g. precipitation, cloud cover, dust) and the system status will continuously be monitored. Using the ETC in realtime, the PH3 will adjust the exposure times accordingly. In addition, each exposure will be reduced automatically after being readout to assess the quality of the observation and adapt the exposure times in realtime. The PH3 will thus run in parallel with Phase 4 (PH4) which governs the automatic data reduction and validation. Figure \ref{fig:phase34} shows an activity diagram for PH3 and PH4.

\subsubsection{Weather conditions and status server}

During PH3, the weather conditions need to be monitored continuously for the safety of the observatory and engineering data related to the status of the observatory will be continuously collected.

The weather conditions under which MSE will operate have been defined based on the historical data recorded at CFHT over the past 10 years. The main weather conditions that can affect science operations are usually the wind speed and the humidity. The current policy at CFHT is to close the dome if the wind speed is above 50 knots or humidity is above 85\%. This usually leads to losses of about 10\% of observing time, or half of all weather losses. Other reasons for weather losses are clouds, for instance. More details about weather losses and observing efficiency are provided in Ref.~\citenum{Flagey2018oe}. There will be weather sensors measuring wind speed, humidity, temperature, dust particles, and others, installed at various places inside and outside the observatory, and the measurements will be monitored in real time from the remote observing room. The measurements will be archived along the data in an Observatory Data Repository (ODR).

The ODR will also contain engineering data, as well as science data. The engineering data are all the telescope and instrument parameters, in addition to voltage, pressure, and various technical data. A status server will monitor and report on these measurements to enable the MSE operations staff to quickly identify issues with a subsystem.

\subsubsection{Technical support of operations}

The purpose of technical operations is to support science operation by ensuring the observatory is technically ready and available for nightly remote observing. Technical operations maintain the performance, reliability and availability of the observatory system by performing predictive (i.e. automatic and autonomous), preventative (i.e. verifications) and corrective (i.e. repairs) maintenance as dictated by the as-delivered observatory.

Technical operations are usually structured based on the preventative maintenance requirements of the observatory systems and subsystems including the frequencies and durations of scheduled maintenance, which define the technical downtime expected. Preventative tasks will be the most common and will follow a pre-defined schedule to service all subsystems in the observatory. Preventative maintenance will rely on a checklist, provided by a preventative maintenance program tool. The goal of the preventative maintenance is to limit the occurrence of unexpected failures that would result in the cancellation of nighttime operations. To limit the time spent on each maintenance task, especially at the summit's elevation, the access to each subsystem will be made as easy as possible.

While the objective of preventative maintenance is to minimize unscheduled technical downtime, predictive maintenance programs remotely monitor the “health” of systems and subsystems at the component-level in order to recognize abnormal operating conditions, impending maintenance issues and potential failures. The goal of predictive maintenance is to act preemptively to eliminate breakdowns that result in unscheduled technical downtime. The philosophy of predictive maintenance is fully compatible with a remote observing approach since the hardware and software monitoring systems required are complementary for these processes.

In addition, technical operations need to accommodate unscheduled technical downtime and perform corrective maintenance, i.e. repairs, due to system and subsystem failures. The unscheduled technical downtime (179 hours per year) is part of the  Observing Efficiency budget presented in Ref.~\citenum{Flagey2018oe}. The corrective maintenance will include repairing subsystems, replacing parts, or modifying a subsystem to prevent additional failures. The technical crew will have access to a clear and exhaustive description of the problem via the ODR and will be able to discuss the problem and the required actions with staff at headquarters.

\subsubsection{Primary mirror maintenance}

At this stage of the MSE project, we have focused on the maintenance of the segmented mirror, which will be a novelty for the operations staff at CFHT. 

It has been common practice for decades to regularly clean the primary mirror of a telescope to maintain a high reflectivity and thus a high throughput of the system. At CFHT, a CO$_2$ cleaning is performed after each instrument exchange, about once every two weeks. At Gemini, a CO$_2$ cleaning is performed weekly but an additional in-situ wash is done every six months to one year (see Ref.~\citenum{Vucina2006}). An in-situ wash implies a standard contact-wash with natural sponges and neutral soap, followed by deionized water rinsing and drying (using small air knives). At the Hobby-Eberly Telescope, a CO$_2$ cleaning that requires two to three hours of work involving three people is performed monthly\footnote{\url{https://het.as.utexas.edu/HET/O_M/Maintenance/mirror_reflec/pmco2clean.htm}}. The primary mirror of MSE will likely require a weekly CO2 cleaning. In-situ wash will not be an option because there is no way to contain the liquid and prevent it from flowing on the electronics of the segment support assembly.

The primary mirror of MSE will comprise 60 segments built with 6 identical sections of 10 unique segments. An additional set of 10 segments will be available as spares. To meet the science sensitivity requirements, the reflectivity of the primary mirror has to be as high as possible between 360 and 1800 nm. MSE will use an enhanced protected silver coating for the primary segments. The MSE primary mirror coating will use a different recipe (ZeCoat protected silver, Ref.~\citenum{Sheikh2016}), which has better blue performances than the Gemini recipe. The Gemini observatories have experienced a loss in reflectivity of at most 3\% after two years at any wavelength between 360 and 1800 nm, based on their protected silver coating (see Ref.~\citenum{Vucina2006}). These losses are larger in the blue (3\% in 2 years at 470 nm) than in the red (1.2\% in 2 years at 880 nm). We assume that each MSE segment will degrade at a maximum rate of 2\% in two years. We will monitor the reflectivity of each segment to schedule the order in which they need to be recoated. The reflectivity of the whole primary mirror will never be allowed to decrease by more than 1\%, below the optimal reflectivity of 60 segments freshly recoated, at any wavelength between 360 and 1800 nm, which means we will stagger the recoating of the segments over two years.

After each segment exchange, a warping, phasing, and alignment of the primary mirror will be necessary. At Keck, this operation takes about 2 hours of nighttime, when things are functioning properly. To minimize the total amount of nighttime lost per year, we will maximize the number of segments exchanged in one operation. At the Keck observatories, a crew of five persons usually replaces three old segments with three freshly coated ones within a day. This does not account for the amount of time required for the recoating process, which we discuss later. At a rate of three segments exchanged simultaneously, and 60 segments in two years, the average period of time between two segments exchange operation will thus be slightly more than one month. The best frequency will be the result of a balance between keeping the schedule simple and regular, maintaining the skills of the staff, dealing with the recoating process, and minimizing impact on observing efficiency. As a baseline, we assume there will be one segment exchange operation per month, 10 months per year, each of them taking care of 3 segments. After each segment exchange operation, the segments that have been removed will be recoated: while it will not be necessary to recoat segments right after they have been removed, a regular monthly schedule will maintain the staff's skills and balanced operations routine.

Because MSE will have a segmented primary mirror with silver coating, Keck and Gemini seem to be good analogs to estimate the time and manpower required to recoat the MSE segments. The segments first need to be prepared (e.g. protection of the edge sensors). At Keck, this is a half-day of work per person and per segment. Before a new coating can be applied, the old coating needs to be stripped. Ideally, at least two segments of MSE will be stripped from their old coating and fully prepared to enter the coating chamber in less than a day. Once stripped, the segments will be recoated as soon as possible. They may then be kept in storage in a safe environment before being installed in the primary mirror. At Gemini (8~m monolithic primary mirror, 150~m$^3$ coating chamber), seven hours are necessary to reach the vacuum level required for their protected silver recipe. In addition, the coating of the four-layer recipe takes about four hours (see Ref.~\citenum{Boccas2004}). At Keck, about five hours are necessary to reach the low-pressure vacuum in the chamber. For MSE, to make the process faster, the coating chamber will be large enough to contain at least two segments at a time and multiple coating chambers might be available to provide redundancy. The pumping and coating operations will happen autonomously, day or night, though care will be taken to avoid peaks in electrical power while other critical operations are occuring. Once the process is complete, each segment's coating will be verified. If the coating does not meet pre-defined specifications, it will have to undergo the whole process of stripping and recoating again.

\subsection{Phase 4}
\label{sec:ph4}

While the data are being observed in PH3, they need to be validated and reduced by the automatic data reduction pipeline (ADRP). The ADRP provides real-time feedback to the OMG to adjust exposure time, if necessary, and thus runs in parallel to PH3 (see section \ref{sec:sched}). The ADRP processes science exposures and generates calibrated science data products. The science data products (raw, reduced, and calibration) will be stored into an Observatory Data Repository (ODR) that will also contain engineering data (system and subsystems parameters) and environmental data (e.g. weather). The user interface to the ODR will help the MSE staff perform quality analysis and troubleshooting, compare calibration methods, test pipeline improvements, and study performance trends. More details about the ADRP are provided in section \ref{sec:adrp}.

\subsection{Phase 5}
\label{sec:ph5}

\begin{figure}
\caption{\label{fig:phase5} Activity diagram for PH5 summarizing the activities related to data distribution and data archive.}
\includegraphics[width=\linewidth]{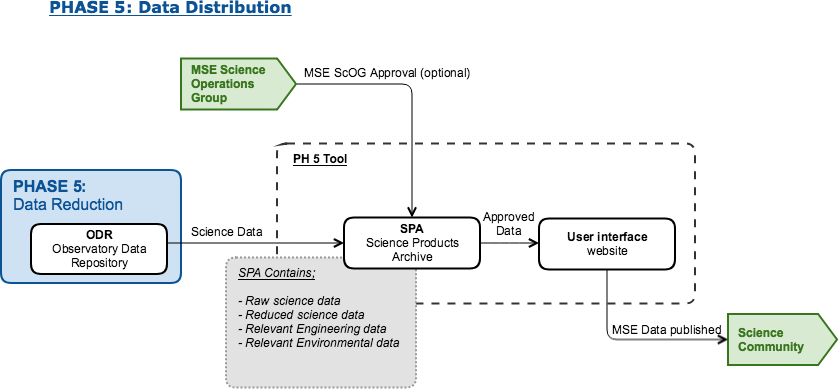}
\end{figure}

Once the data have been reduced in PH4, they can then be distributed to the survey teams, which constitutes the Phase 5 (PH5). Whereas the PH4 products are aimed at the observatory staff for quality analysis and troubleshooting, the PH5 products are aimed at the Science team. Most engineering logs will therefore not be archived into the Science Products Archive (SPA) but only in the Observatory Data Repository (ODR). The user interface to this database will likely be similar to archive websites provided at other astronomical data center. Figure \ref{fig:phase5} shows an activity diagram for PH5. More details about the data distribution are provided in section \ref{sec:adrp}.

\section{Scheduling}
\label{sec:sched}

In this section we describe the process by which observations with MSE are defined, scheduled and queued for observations. We discuss how complex an observation can be, what information it will contain, and how it will be executed.

\subsection{Definitions}

\begin{figure}
\centering
\caption{\label{fig:OS} Diagram showing examples of envisioned sequence of exposures for MSE's low-resolution spectrographs.}
\includegraphics[width=.75\linewidth]{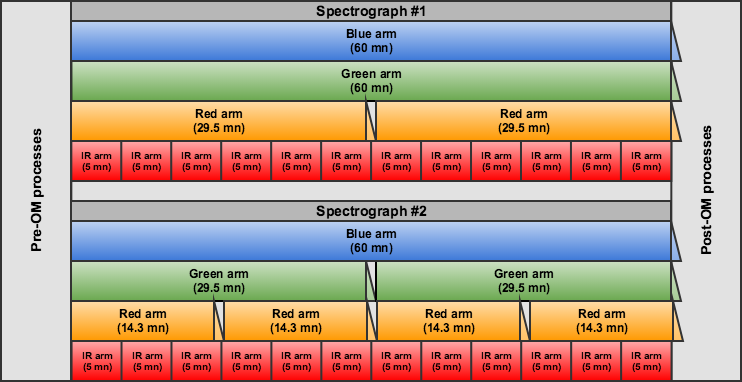}
\end{figure}

An {\bf Observing Sequence (OS)} describes the collection of the numbers and durations of all exposures for all arms of all spectrographs. An exposure is the time during which a detector collects photons, until readout. Figure \ref{fig:OS} shows what an OS could be for two LR spectrograph units.

An {\bf Observing Field (OF)} corresponds to a collection of Observing Elements (OE), being allocated to fibers within one field of view of MSE. An OF is selected and built from the content of the OED. In addition, an OF contains positions for sky background and calibration stars measurements. Figure \ref{fig:OF} shows what an OF could be with allocated targets to the 1083 HR fibers.

An {\bf Observing Matrix (OM)} is the combination of an OF and an OS, created by the Observing Matrix Generator (OMG), to form a unique block of observations. An OM combines the targets locations in the sky (OF) with the exposure times for each arm of each spectrograph (OS).

In their simplest forms, an OF would use OEs from a single survey program and an OS would use the exact same exposure times for all spectrographs and all arms. However, more flexibility is anticipated when defining OMs for MSE, because of its significant multiplexing capabilities, its broad spectral coverage, and its suite of multiple spectrographs (multiple LMR units and multiple HR units), each with multiple arms. This flexibility will improve the utility of the science data gathered by MSE. It is also expected that for most surveys, more than one OM will be required before observations in a particular area of the sky are declared complete.

\subsection{Observing Sequence}

There are two essential and related properties to keep in mind about an OS, to keep it both simple and versatile. First, within an OS, different exposure time sequences on each spectrograph and different exposure times on each arm within a spectrograph can be used. Second, an OS, and thus an OM, will stop when the longest exposure from any arm of any spectrograph is read.

Different exposures times on each spectrograph and arm will help mitigate readout noise, saturation and non-linearity issues. The goal is to optimize the signal to noise ratio (SNR) of the targets that make up the OF. This means taking into account multiple effects, including but not limited to: (1) the sky brightness which will often be the main source of photons, (2) the well depth of the detectors that will be filled faster in some wavelength regimes than others, (3) the cosmic ray hits that may require multiple repeats of a given exposure to mitigate their impact, and (4) the atmospheric differential refraction (ADR) which will make targets apparently drift across the field of view and thus decrease the injection efficiency into the fibers at the focal plane (see Ref~\citenum{Flagey2018ie} for more on the injection efficiency).

\begin{figure}
\centering
\caption{\label{fig:seq} Diagram showing the envisioned nighttime sequence of events for MSE with allocated times.}
\includegraphics[width=\linewidth]{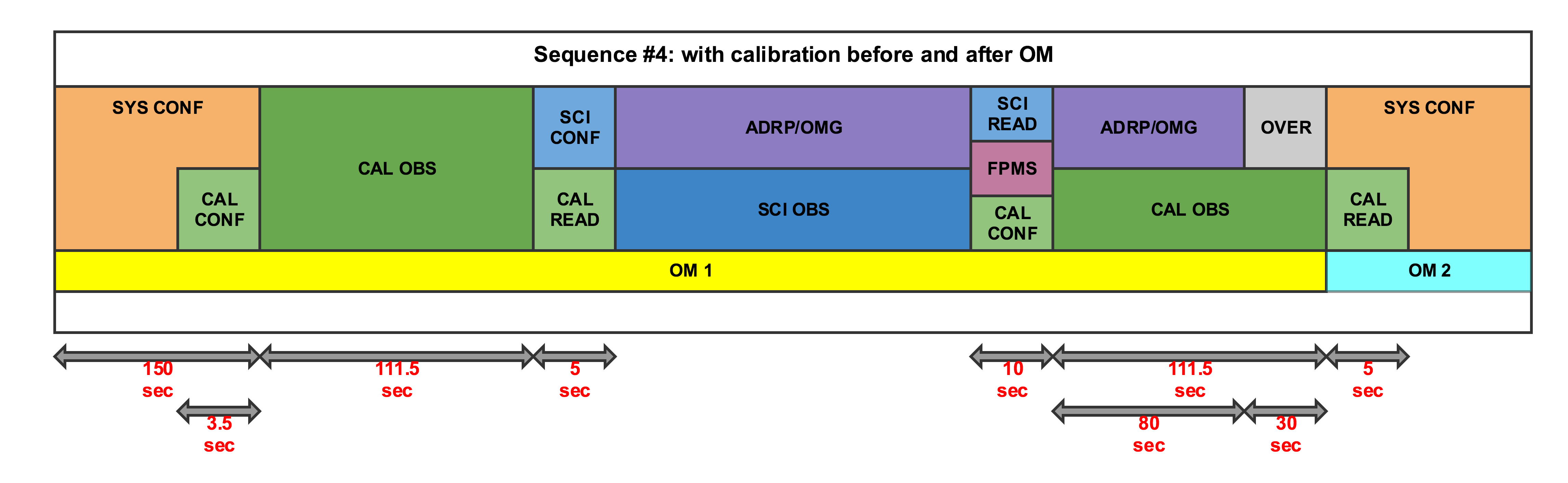}
\end{figure}

Within our current understanding of the MSE system, we expect that exposures in the NIR will typically be about 5-minute long while exposures in the blue could last up to one hour. Because the NIR and blue arms will be in the same spectrograph, and to maximize the SNR in the blue arm, there will be multiple short exposures in the NIR in parallel with a single long exposure in the blue (see Figure \ref{fig:OS}). Once the longest exposure has been complete, the OM will stop. Evidently, longer OMs will increase the observing efficiency of the observatory (i.e. the fraction of time not lost to weather that is spent collecting photons). In Ref~\citenum{Flagey2018oe} we show that, with the adopted baseline for nighttime calibrations (see section \ref{sec:calib} and Figure \ref{fig:seq}), if the average length of an OM exceeds 44 minutes, MSE will meet the Observing Efficiency science requirement of 80\%.

\subsection{Observing Field}

The first degree of complexity in an OF will come from the mixing of Observing Elements (OE) from different observing programs, and thus mixing OEs that use the LMR spectrographs with OEs that use the HR spectrographs, which is possible thanks to the Sphinx positioners design (see Ref~\citenum{Smedley2018}). We assume however that most observing programs will use either the LMR or the HR spectrographs, and only some programs will use both. Consequently, we expect that the selected observing programs might not provide both LMR and HR targets for every single field that MSE will observe. Therefore, the Science team and ScOG will provide targets in addition to those provided by the observing programs as “filler” targets. These additional targets will make up a pool of OEs with a low priority in the OED. There will be an automatic way to identify the regions in the sky where there is a deficiency of LMR or HR targets.

In addition to fibers allocated to targets from the Survey programs, a number of fibers will be allocated to sky positions and calibration stars, for data reduction and calibration purposes (see sections \ref{sec:adrp} and \ref{sec:calib}). We anticipate that at least 10\% of the fibers will be allocated to sky positions while a few tens of fibers will be allocated to calibration stars. Calibration stars will be listed in the OED and defined by the MSE ScOG thanks to the PH2 Tools. For the sky position, a catalog will be built as a collection of positions away from sources identified in external databases. The sky fibers and calibration star fibers will be distributed as homogeneously as possible among spectrographs to account for calibration variations between spectrograph units, and across the field of view to properly account for the spatial variations of the sky brightness and transmission.

\subsection{Automatic scheduling}

The selection of an OF and the selection of an OS will not be done manually due to the number of targets in the OED and the number of fibers in the field of view. The Observing Matrix Generator (OMG) will combine the information contained in the OED, the current and expected observing conditions, the current and possible instrument configurations, the time available for science, and multiple other constraints to schedule OMs, i.e. define the best next OM and the best sequence of OMs. The OMG will also be responsible to queue the next OM, i.e. send commands to the observatory so that the OM can be executed. Finally, the OMG will also update in real-time the current OM and the upcoming sequence of OMs based on feedback received from the ADRP, the weather sensors, and other system status. More details about the OMG are given in section \ref{sec:omg}.

\begin{figure}
\centering
\caption{\label{fig:OF} Illustration of the target/fiber allocation process. The HR Sphinx positioners individual patrol regions are shown as red disks. Targets (grey crosses) are provided over a slightly larger field of view and only those marked by a black cross can be reached. In the zoomed panel, red lines and crosses show which target is allocated to which fiber. The bottom panels show the allocation after multiple pass on the same field where the previously allocated targets are shown as green crosses.}
\includegraphics[width=.45\linewidth]{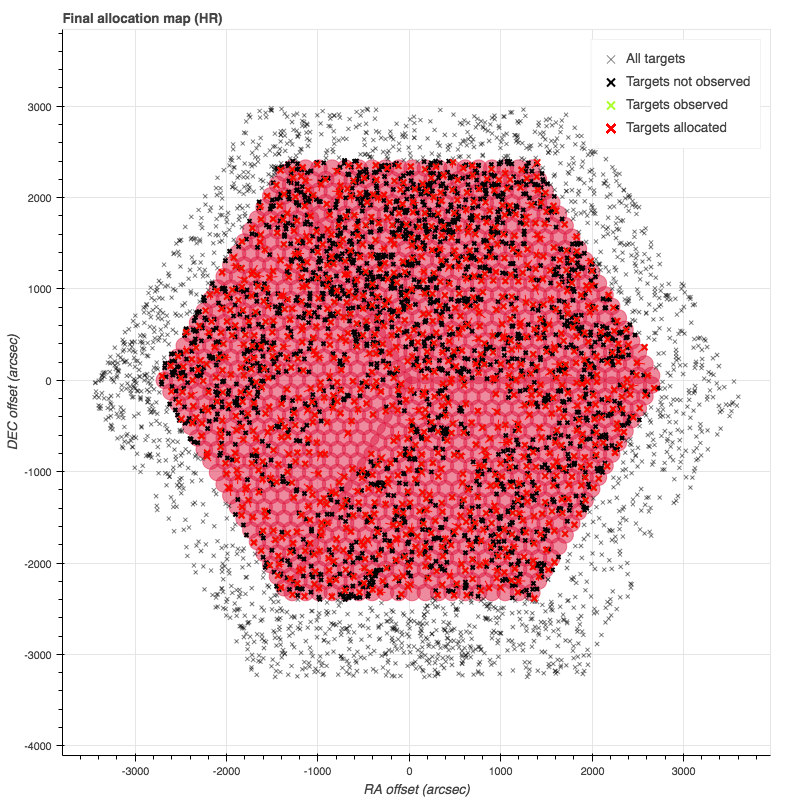}
\includegraphics[width=.45\linewidth]{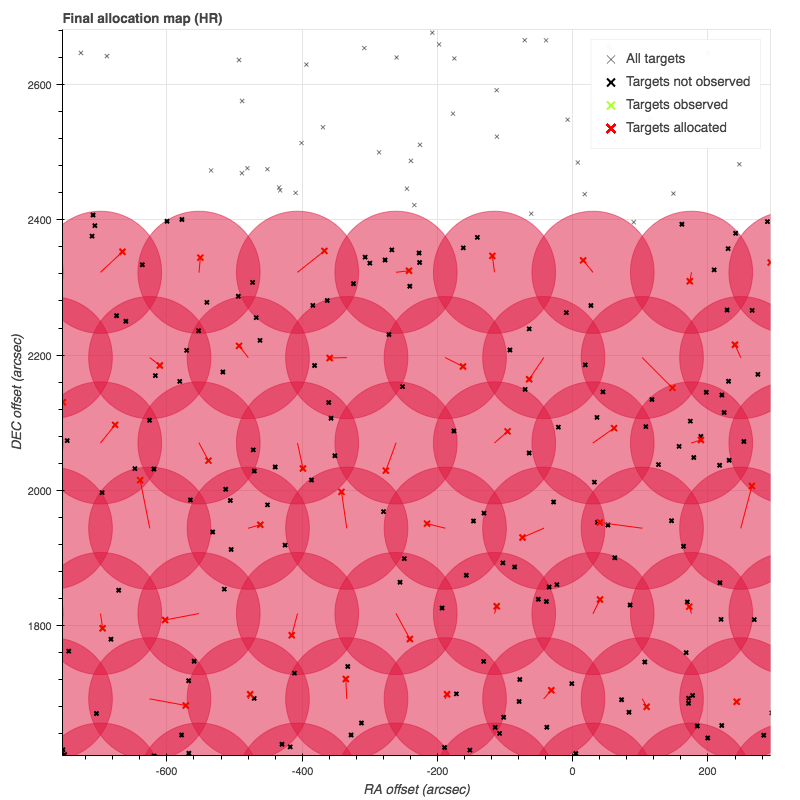}
\includegraphics[width=.45\linewidth]{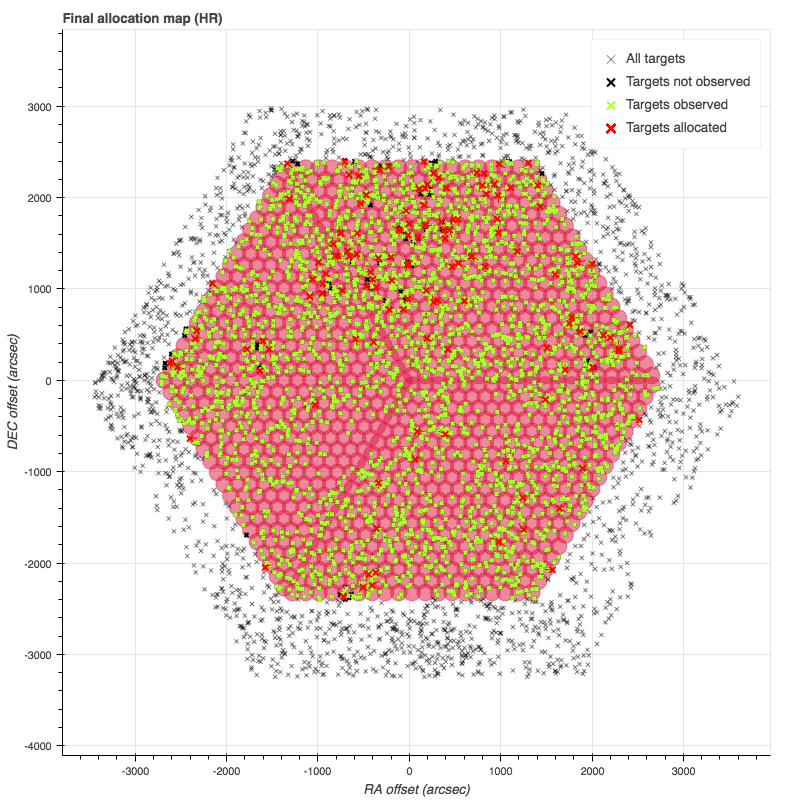}
\includegraphics[width=.45\linewidth]{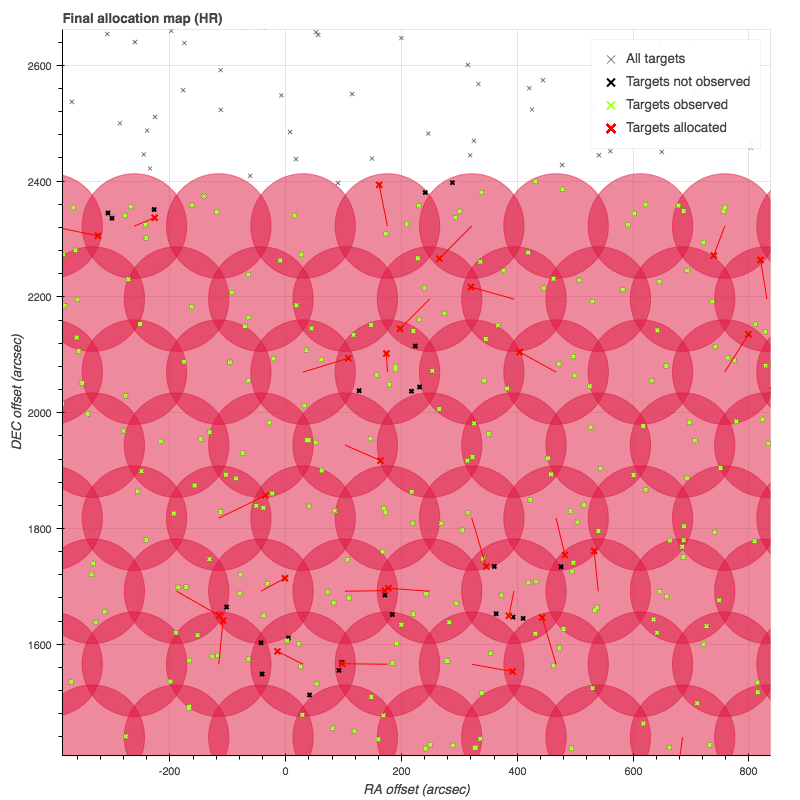}
\end{figure}

\section{Tools}
\label{sec:tools}  

In this section we present the main pieces of software that are envisioned to support the operations of MSE. Some, like the Exposure Time Calculator (ETC) or the Target Allocation Simulator (TAS), already have a version in development. Others, like the Observing Matrix Generator (OMG) and the Automatic Data Reduction Pipeline (ADRP) are still at the conceptual phase.

\subsection{Exposure time calculator}
\label{sec:etc}

\begin{figure}[h]
\caption{\label{fig:etc} Screenshot of the current input parameters table for the ETC for MSE.}
\includegraphics[width=\linewidth]{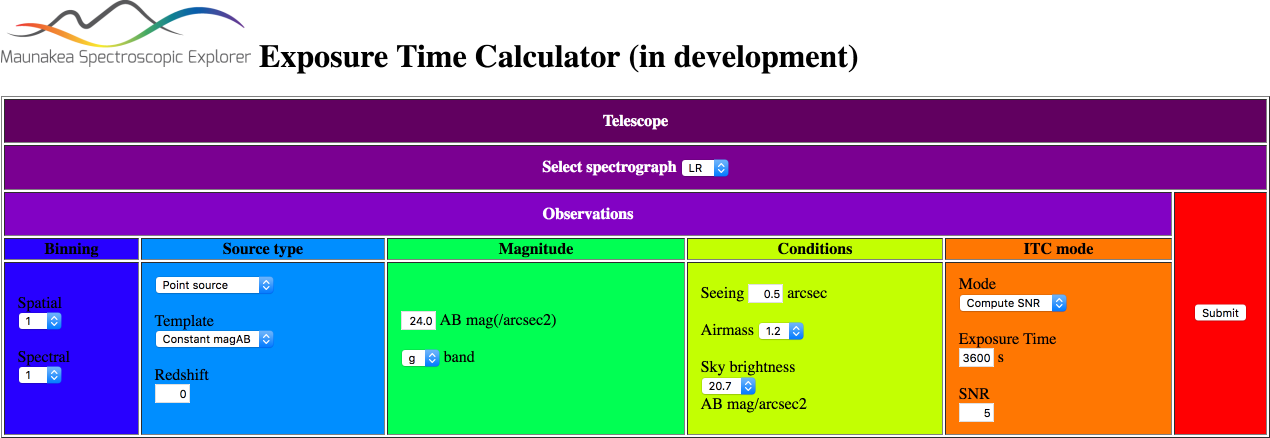}
\end{figure}

The ETC for MSE is a tool that will support the scientific community willing to submit observing program by simulating the sensitivity of MSE under various conditions. Figure \ref{fig:etc} shows the current user interface to the ETC. The user can chose the spectrograph (LR, MR, or HR), define the source via templates, and the observing conditions. The ETC can be found at {\url{etc-dev.cfht.hawaii.edu/mse/index.html}}. In its current version, the ETC for MSE can either provide the SNR at all observed wavelengths given an exposure time, or provide the time required to reach a given SNR at all wavelengths. The ETC relies heavily on the systems budgets for throughput, noise, and injection efficiency (see Ref.~\citenum{Flagey2018ie}, \citenum{McConnachie2018b}, and \citenum{Hill2018} for more details).

The ETC will also be used during operations to predict the time required to reach a given SNR on a given target under the actual conditions experienced at the summit. The results from the ETC will be compared to those actually measured on the collected data to assess the quality of the data and possibly identify biases or issues in the system. Sections \ref{sec:omg} and \ref{sec:adrp} provide additional details.

\begin{figure}[h]
\caption{\label{fig:etc_res} Example of results from the ETC for MSE. In this case, the SNR is shown for a target of constant magnitude $mAB=22$ in the five low-resolution bands: three optical bands, and both $J$ and $H$ bands.}
\includegraphics[width=.5\linewidth]{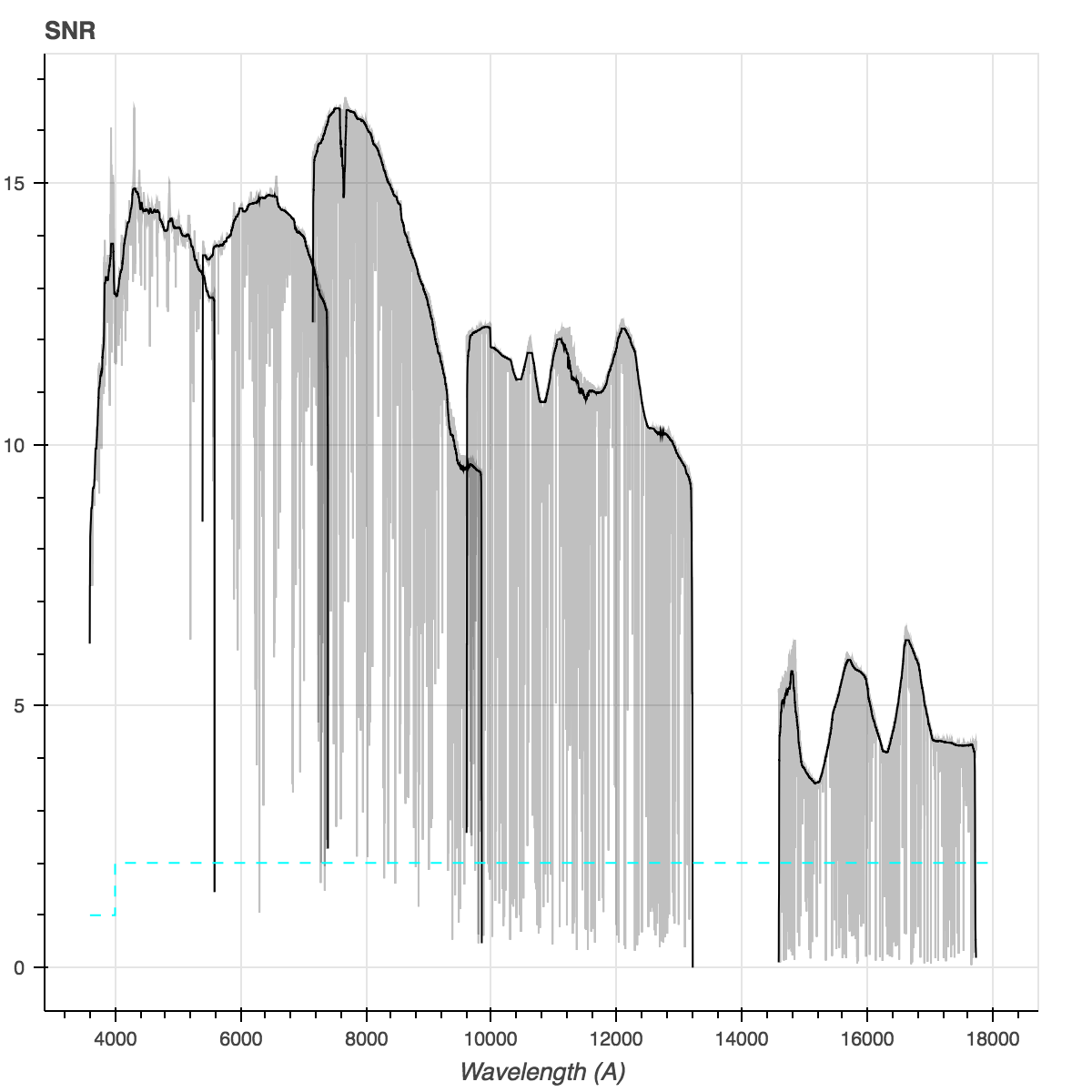}
\end{figure}

\subsection{Target Allocation Simulator}
\label{sec:tas}

The TAS goal is to allocate as many fiber/target pairs as possible by minimizing the total fiber-to-target distance. In an ideal case, all fibers can be allocated to their nearest target and all targets to their nearest fiber. When targets or fibers are competing for their nearest possible allocation, five different allocations sequences are tested, with the fibers or targets that are competing chosen in a random order and allocated in turn to their nearest fiber/target. The sequence out of the five that leads to the lowest total fiber-to-target distance is then chosen.

The TAS can be found at \url{etc-dev.cfht.hawaii.edu/mse/alloc.html}. It allows the user to provide a list of targets for either the LMR or the HR spectrographs, where each targets has a Priority and a Repeat ranking. The user can define the center of the field of view or let the TAS do it. The user then choses whether they want to find out (method 1) the fraction of allocated targets after a given number of iterations or (method 2) how many iterations are needed to allocate a given fraction of the target list.

After a complete allocation has been performed, the TAS decreases the Repeat ranking of every allocated targets by one. If multiple allocations are necessary, the targets that are kept in the list are those with a Repeat ranking of at least 1. Targets with a higher Priority ranking have more chances of being allocated. For the purpose of computing the total fiber-to-target distance, the TAS considers that the distances between a fiber and all targets within its patrol region are the true distances divided by the Priority ranking of each target. For instance, a Priority 4 target 40\arcsec\ away from a positioner is actually assumed to be 10\arcsec\ away when the TAS minimizes the total fiber-to-target distance.

Figure \ref{fig:OF} shows an example of the target/fiber allocation process for the HR positioners. The targets in the field were selected from the SDSS DR9 catalog near $(RA, DEC) = (30, +62)$ and only targets with a $u$-band magnitude between 19.5 and 20 were kept. In this example, 3032 targets can be reached by an HR positioner out of the 4878 targets provided by the user and surrounding the field of view. There are 1026 HR fibers that can reach a target out of the 1083 HR fibers in the field of view. Working with those, the TAS found 1019 pairs on the first iteration. The numbers of allocations after each iteration are shown in Table \ref{tab:tas}. From one simulation to another, using the same target list and input parameters, we noticed that the results are varying by less than 0.1\% which characterize the great repeatability of our method.

\begin{center}
\begin{figure}[h]
\caption{\label{fig:tas} Screenshot of the current TAS webpage for MSE.}
\includegraphics[width=\linewidth]{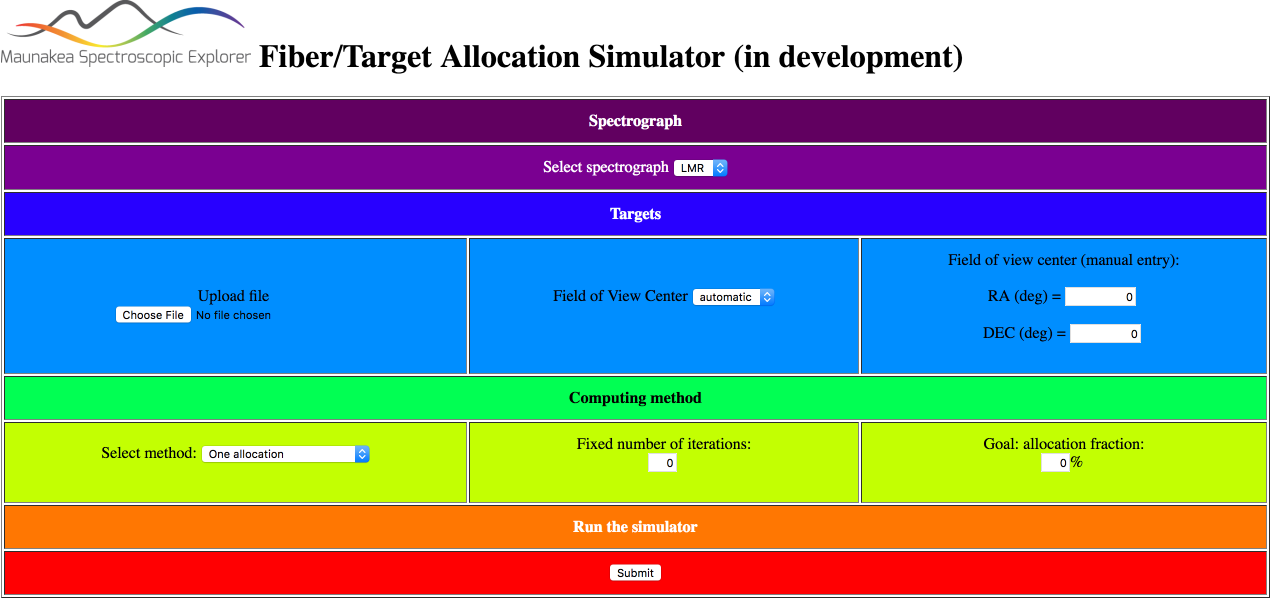}
\end{figure}
\end{center}

\begin{table}[h]
\centering
\caption{\label{tab:tas} Typical results from the TAS corresponding to the simulation presented in Figure \ref{fig:tas}. The allocated fraction is the cumulative number.}
\begin{tabular}{| l r r r r |}
\hline
Iteration & Targets left & Fibers in reach & Pairs allocated & Allocated fraction\\
\hline
1 & 3032 & 1026 & 1019 & 33.6 \\
2 & 2013 &  884 &  864 & 62.1 \\
3 & 1149 &  649 &  609 & 82.5 \\
4 &  540 &  369 &  332 & 93.3 \\
5 &  208 &  158 &  138 & 97.7 \\
6 &   70 &   50 &   44 & 99.1 \\
7 &   26 &   20 &   17 & 99.7 \\
8 &    9 &    7 &    7 & 99.9 \\
9 &    2 &    2 &    2 & 100 \\
\hline
\end{tabular}
\end{table}

The limitations of the TAS at the time this paper is submitted are:
\begin{itemize}
\item no fibers are allocated to sky or calibration star position,
\item all fibers and positioners are working (i.e. there are no technical failures),
\item there is no dithering and the field of view remains in a fix position for all iterations, and
\item there is no lower limit to the distance between two fibers (the Sphinx design recommends to keep fibers 1.0~mm or 9.4\arcsec\ apart).
\end{itemize}

\subsection{Observing Matrix Generator}
\label{sec:omg}

The Observing Matrix Generator (OMG) will be the scheduler for MSE. It will define the best field to observe at any given time, given the current and expected weather conditions (e.g. cloud coverage, Moon illumination), given the current and expected system status (e.g. telescope current pointing). In addition to sending all required information to the observatory to observe the next field, the OMG will also generate a schedule for a longer period of time (e.g. upcoming night to upcoming year). This will enable the MSE operations staff to estimate the typical observing efficiency based on historical weather and engineering data.

The OMG will generate an observing matrix (OM), i.e. the combination of an observing field (OF) and an observing sequence (OS). Details about OM, OF, and OS can be found in section \ref{sec:sched}. The OMG will define the best collection of targets from the OED that can be observed simultaneously within one field of view of MSE, given the positioners patrol region and the time required to observe each target. The optimization will rely on many parameters (e.g. program's priority, target's priority) and will combine aspects from an imaging survey scheduler with a target/fiber allocation simulator.

An imaging survey typically only needs to optimize the mosaicing or tiling of a given area of the sky and reach a given depth as homogeneously as possible over that area. Once a given field of view has reached that depth, every pixel within that field of view is considered observed and the scheduler will not have to come back to any of those pixels. An imaging survey scheduler basically deals with a one-dimension parameter space defined by fields of view coordinates on the sky sphere.

A spectroscopic survey like MSE not only needs a scheduler that optimizes the field of view position but also needs to consider the optimization within each field of view, which depends on the exact coordinates of the field of view. This is because the size of the positioners patrol regions is 9.63~mm or about 90\arcsec\ on the sky (7.77~mm pitch $\times$ 1.24). The sampling for optimizing the field of view position might thus need to be much finer than for an imaging survey. We envision that some simplifications might help the OMG generate the schedule in real-time (see below). For instance, scores could be attributed ahead of time to a coarse grid of sky coordinates using all the information in the OED, based for instance on the local density of targets, and the OMG would then focus on the high-score areas only.

A critical piece of information not provided by the observing programs will be the location of sky positions, calibration stars, and guide stars. The MSE ScOG will be responsible to generate lists of specific OEs with the PH2 Tool for sky positions, calibration stars, and guide stars. Guide stars will be bright enough to provide accurate guiding and at the same time numerous enough to appear in each of the three acquisition and guiding cameras (AGC) in the focal surface of a given OM. A finite list of available pointings for MSE might be derived from the positions of guide stars, if their typical density is low. Calibration stars will be used to correct for telluric absorption and perfom photometric calibration. To avoid significant cross-talk between faint targets and typically bright standard stars, catalogs of fainter calibration stars will be established before the start of operations for MSE. The sky positions will be generated using deep imaging surveys and defined as regions away from any potential contaminating source. All these positions (sky, calibration, guide) will be added in the OED and labeled accordingly. In each OM, the sky fibers and calibration stars fibers will be distributed as homogeneously as possible among spectrographs to account for calibration variations between spectrograph units, and across the field of view to properly account for the spatial variations of the calibration.

Once an OM is generated, it will be sent to the observatory control sequencer so all subsystems can be configured to start the observations. The OMG will queue one OM at a time and keep the sequence of upcoming OM in the background. In addition, while an OM is being observed, the OMG will take into account the current weather conditions and real-time feedback from the automatic data reduction pipeline (ADRP, see section \ref{sec:adrp}) to adjust the OS (i.e. exposure times and number of exposures) for the current OM and adjust the sequence of upcoming OMs. For instance, the OMG will reduce the length of an exposure or decrease the number of exposures if the seeing has improved since the beginning of the OM, or increase them if the seeing has worsened. These adjustable exposure times will allow the OMG to respond to varying observing conditions during an OM, and given the long exposures expected in an OM, to improve the observing efficiency. The OMG will need to have access to enough information while the OM is executed to make those decisions. In particular, the OMG will have access to all weather sensors information stored in the ODR, the predictions from the ETC, and the results from the ADRP.

\subsection{Automatic Data Reduction Pipeline}
\label{sec:adrp}

The MSE observatory will produce a very large amount of data: raw/reduced, science and calibration data, engineering subsystems parameters logs. Some of these data will be released into an archive for access by scientists, while others will be needed mostly for quality analysis and troubleshooting at the observatory. Therefore, MSE will have a database and a user interface for PH4 (data reduction) that may slightly differ from the database and user interface for PH5 (data distribution). The Automatic Data Reduction Pipeline (ADRP) will be responsible for the data reduction and data validation during PH4. The ADRP will also provide realtime feedback to the OMG so the schedule can be updated if necessary.

In PH4, the ADRP will automatically reduce raw data. After an exposure is readout while an Observing Matrix (OM) is observed in PH3 (nighttime operations), the ADRP will automatically reduce it and update the Observing Element Database (OED) to confirm whether each OE in that OM requires more observing time or not and trigger updates of the schedule by the OMG (see section \ref{sec:omg}). Raw data, reduced data, and calibration data will then be added to the Observatory Data Repository (ODR), along with all system and subsystems engineering logs, and weather logs. The science data (raw and reduced) will remain in the ODR to allow the observatory staff to perform quality analysis, comparative analysis (e.g. between different calibration methods or different ADRP versions), and establish historical trends based on science data. The user interface to the ODR will be developed to support operations at MSE, facilitating queries and downloads of data obtained on a given night or under certain conditions. The ODR will contain individual tables of data (science, engineering, environmental) that should then be easily cross-matched to facilitate this analysis work. The ODR will also serve as database for the system status server used in the context of the observatory operations. We envision that the ADRP will be the process in charge of populating the ODR with science data, while another process will be responsible to archive engineering and environmental data into the ODR.

The database in PH5 will be the Science Products Archive (SPA), which will contain all science data and only the relevant engineering and environmental data to be distributed to the Science team, likely in the form of metadata saved in the header of the files in the SPA. The user interface will likely take the form of a website, similar to what other astronomical data centers are providing to their users, to facilitate the query and download of specific targets or survey program as well as data mining analyses. While the ADRP will be responsible to automatically reduce the data in PH4, another process will likely be taking care of the data distribution, making sure in particular that the ODR and SPA are properly synchronized. That process will be made as automatic and autonomous as possible, although it is likely to require MSE staff's approval before releasing data products outside the observatory.

\subsubsection{Science products}

MSE will provide spectroscopic data of astrophysical sources at three different spectral resolutions. Both types of spectrographs will always be available for use, and hence typically more than 4300 sources (science objects, sky, or calibration sources) will be observed simultaneously. To illustrate the expected size of the dataset, given the conceptual designs of the HR and LMR spectrographs, which use three and four arms respectively, if we adopt an average exposure time of 30 minutes and an average night length of 8 hours, then this implies MSE could routinely produce close to 200,000 individual spectra per night and more than 60 million per year. These individual spectra, one per spectrograph arm, per exposure, per object, form the main data product of the MSE system. Different levels of data products are then anticipated for MSE.

\begin{description}
\item[Level 0] data products are the raw data, unprocessed readouts from the detectors, for science images and nighttime/daytime calibration images.
\item[Level 1] data products are the first version of the dark/bias corrected, flat-fielded, 2D images, and wavelength calibrated, flux calibrated, sky subtracted 1D spectra that are generated soon after the end of a science exposure at night. Level 1 data products will not necessarily use optimal calibration data and recipes since these may not yet be available (e.g. daytime calibration will be obtained during the following day, see section \ref{sec:calib}). Level 1 data products are mostly intended for quality assurance by the MSE staff and initial science analysis by the Survey team.
\item[Level 2] data products are 2D images and 1D spectra produced by MSE after all calibrations (nighttime and daytime) have been obtained. Level 2 data products will use the best possible calibration exposures and reduction recipes. Level 2 data products are the main homogeneous, science ready data products generated by the MSE observatory including both individual-epoch spectra and co-added spectra combining different observations at the same resolution. A basic set of science parameters will be included in Level 2 data products.
\item[Level 3] data products will be any processed images and spectra, derived object parameters, and associated metadata provided by the Survey teams. They will take the form of catalog releases at regular intervals. Each Survey team is expected to have their own specialized analysis methods and major science products, and thus while these data products will be homogeneous within a survey, they will be different between surveys and from the MSE Level 2 releases. It is likely, however, that the MSE partnership will require certain core products to be produced by all surveys. Level 3 data products will be available in the SPA and likely not in the ODR.
\item[Level 4] data products will be images, spectra, and/or associated metadata provided by the broader Science team and/or international community, and are considered “value added” data products. Level 4 data products will be available in the SPA and likely not in the ODR.
\end{description}

\begin{figure}
\caption{\label{fig:sciprod} The various levels of Science products for MSE, with indication of the expected content, the group responsible for producing the products, the envisioned access rights, and whether multiple releases are anticipated.}
\includegraphics[width=\linewidth]{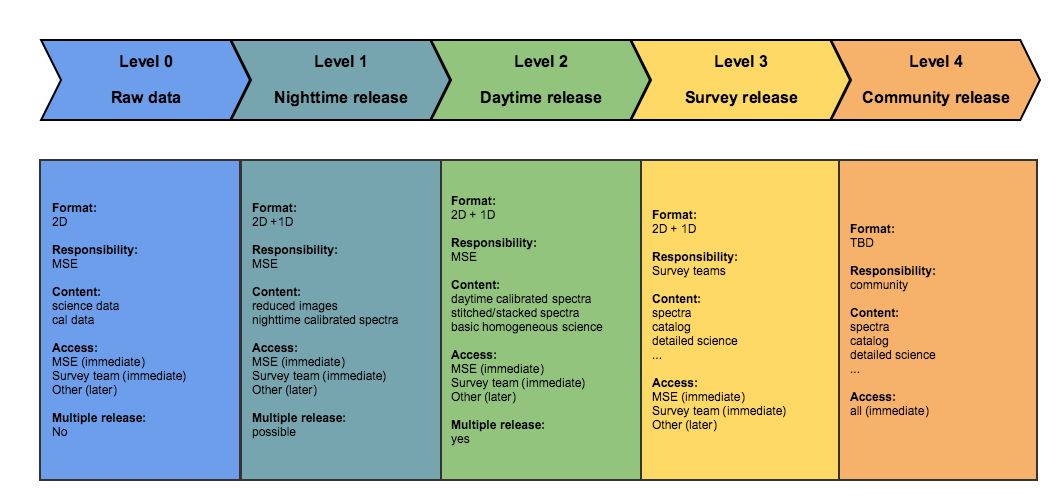}
\end{figure}

MSE will collect thousands of spectra every hour. The goal of PH4 is to reduce these data and validate them so they can be distributed in PH5. The reduction and validation process described below only concerns the data products that are under the responsibility of MSE: Level 0 to 2. The distribution process also comprises Level 3 and 4.

\subsubsection{Data reduction}

The ADRP will be executed automatically multiple times for a given OM:
\begin{description}
\item[After each readout at night:] an OM will correspond to multiple readouts, thus the ADRP will be executed multiple times during an OM. This first data reduction will be used for real-time feedback on the data and enable the OMG to adjust the schedule (see section \ref{sec:omg}). The data products generated at that time are the Level 1 data products.
\item[During the day:] once all daytime calibrations relevant to the science observations have been obtained (e.g. daytime flats and arcs), the ADRP will generate some of the Level 2 data products. We envision that this will happen at least once a week. For more about calibrations, see section \ref{sec:calib} and Ref.~\citenum{McConnachie2018a}.
\item[For repeated OE:] after an OM which contains an OE that has been observed in a previous OM, the ADRP will generate some of the advanced Level 2 data products (e.g. stacked spectra).
\end{description}

\subsubsection{Data validation}

Each time a spectrum has been reduced by the ADRP, it will go through a quality analysis (QA) process. This will correspond to the “data validation” at the observatory and will be the main responsibility of the ScOG. The objective of the ADRP in terms of QA is to automatically associate flags to each spectrum that indicate its quality and its utility for science measurements.

The QA will be based on the data themselves, using in particular the estimation of the signal-to-noise ratio (SNR) derived by the ADRP, on the engineering and environmental data. A significant discrepancy between the expected SNR (via the ETC, see section \ref{sec:etc}) and the measured SNR will be flagged. The exact nature of the flag will depend on the reason(s) for the discrepancy, which will likely be found in engineering and environmental information. The ADRP will need access to weather conditions, either via the weather sensors or via measurements made using the Acquisition and Guiding Cameras (AGC). The ADRP will thus be able to provide an estimate of the sky background, extinction, and image quality during an OM. Those results will be used in the QA process and saved into the ODR and SPA.

The ADRP will automatically generate statistics and plots for the MSE staff. For instance, the ADRP may create a plot of the measured SNR as a function of the expected SNR by the ETC, color coded with the target's magnitude or positioner tilt angle, for all OE in a given exposure, or a map of the discrepancy between expected and measured SNR across the field of view. These will help revealing biases related to the observatory operations.

The QA will occur each time the ADRP is executed. Therefore, Level 0, Level 1 and Level 2 data products may have a different list of quality flags, covering issues related to the execution of the OM (e.g. variable seeing, clouds), the raw data (e.g. cosmic rays, bad pixels), and the reduction pipeline (e.g. issue with flat or dark). The flags will be saved with the products into the ODR and SPA since they will be relevant to both operations and science.

As part of the information comprised in the QA flag, a “validation” flag will be provided for each spectrum. This validation flag will necessarily be binary and indicate whether a given spectrum is considered useful for science or not. This validation flag will be automatically derived from all other quality flags and some combination rules, which may depend on the survey program. This verification process will be supervised by the MSE staff. The ScOG may filter the data and focus on those that have encountered issues, thus facilitating the implementation of diagnostics.

The quality and validation flags for each spectrum will be related to a “success metric”, likely the SNR measured by the ADRP. Because Level 1 and 2 data products will be under the responsibility of MSE, the metric will ideally use a unique definition, consistent with that used by the ETC in PH2, to ensure homogeneity. The SOG and ScOG will work with the Survey team to develop the ADRP and incorporate more complex definitions of the metric if necessary. They will also work closely to provide additional scientific value in the Level 2 data products.

At Level 1 and 2, the ADRP will provide the metric for each individual piece of spectrum obtained for a given OM, as part of the associated metadata. At the Level 2, the ADRP will also provide the metric for the stitched, stacked, and/or co-added spectra. The metric values will be saved into the ODR, the SPA, and the OED to estimate if an OE has been completed.

\subsubsection{Data completion}

Combining all spectra obtained for a given OE and their “validation” flags will lead to a “status” flag for that OE in the OED. The status flag will have at least three different values: completed (the OE has been sufficiently observed to meet the requirements from the Survey team), started (the OE has already been observed but does not meet yet the requirements from the Survey team), or unobserved (the OE has not been observed yet) which is the default value for any OE entered in PH2 (see section \ref{sec:ph2}). The status flag may also indicate a “completion” rate for each OE, as the ratio between the observed SNR and the required SNR, or between the number of visits executed and the number of visits required. Other definitions will be developed as necessary.

The status flag is a critical piece of information for the OMG to perform properly. Completed OEs will by default not be scheduled anymore. Started OEs will likely receive a higher priority, to provide a higher completion rate for the program it belongs to. Ideally, MSE will observe each OE until it has reached the requirements entered in PH2 (e.g. SNR goal or number of visits), and then flag that OE as “completed” in the OED. In reality, some flexibility will be allowed, in agreement with the priorities defined by the Science team.

\subsubsection{Data distribution}

The ADRP will automatically add data products to the ODR and SPA. The MSE staff and the Survey team will have access to all MSE data, although we expect there might be a delay between the time some data products are available to both groups, as the MSE staff first needs to validate the data.

Level 0 and Level 1 data products will immediately be saved into the Observatory Data Repository (ODR) and Science Products Archive (SPA) after each readout to become accessible to the MSE staff for quality analysis and to the Survey teams for preliminary science analysis. Level 2 data products will immediately be saved into the Observatory Data Repository (ODR) after the ADRP has generated them to become accessible to the MSE staff for quality analysis. They might be released into the SPA only after the MSE staff has approved them.

The goal is to have the Level 1 data products ready for the Survey team, the ScOG and SOG during the night they have been obtained and at the latest for quality analysis (QA) during the following day. The Level 2 data products, because they will need additional daytime calibrations, will be available to the ScOG and SOG within a day, and because they may require a more thorough QA, may be available to the Survey team a few days later. Ideally, the ODR and SPA will be linked to the OED. For instance, a user querying the OED will be able to quickly access the data products for that OE in the ODR and SPA, assuming they have permissions to do so. MSE will be responsible for the user interface (e.g. website) to the OED, ODR, and SPA.

Points of contact will be notified automatically when data products become available to enable short term feedback from the Survey teams to the operations groups and allow the Survey teams to analyze and publish MSE data as fast as possible. This will be particularly critical in the early phase of MSE.

\section{Observing efficiency}
\label{sec:obseff}

Among the science requirements that primarily relates to operations, there is the need to reach an observing efficiency of 80\%, where the observing efficiency is defined as the amount of nighttime spent collecting science photons on sky divided by the amount of nighttime not lost to weather.

Given the nominal nighttime sequence shown earlier in this paper (see Figure \ref{fig:seq}), given information received through Conceptual Design Phase on the numerous subsystems in MSE, and given historical data on various technical losses, we established a budget for the observing efficiency, which is described in Ref.~\citenum{Flagey2018oe}. In summary, MSE will reach an observing efficiency of 80\% if the allocations in the budget are all met and if the average duration of a science block of observations is at least 44 minutes.

Allocations in the observation efficiency budget will flow down to reliability, maintenance, and performance requirements on subsystems as well as to an operations plan that will be developed at a later stage.

\section{Calibration}
\label{sec:calib}

Besides the requirement on the observing efficiency, several calibration requirements flow down from the Science to the Operations. These are the requirements on sky subtraction (line and continuum), on velocity accuracy, and on spectrophotometry.

To meet those requirements, we have developed a calibration plan which describes overall principles that MSE will follow. Our principles are that (1) the calibration exposures should provide high SNR frames, (2) the calibration sequence should use as little nighttime as possible, and (3) the calibration photons should accurately reproduce the light path of the science photons. At that stage in the project, we have decided to plan for the worst and hope for the best, making sure the calibration plan we have established will provide the data quality defined by the science requirements on sky subtraction, velocity accuracy, and spectrophotometry. More details are provided in Ref.~\citenum{McConnachie2018b}.

\section{Conclusions}

In this paper we presented the operations concept for MSE, a facility dedicated to highly multiplexed, spectroscopic survey of the faint universe in the optical and near-infrared, which will replace CFHT. The data for MSE will follow the usual phases of operations in an astronomical observatory: submission and selection of proposals, definition of the targets' samples, execution of the observations, reduction and validation of the data, distribution of the data. The operations at MSE will borrow from the operations at observatories in queue service mode and will rely on the decade-long experience of the operations staff at CFHT. However, MSE will have specific characteristics not shared with CFHT (e.g. the number of targets to deal with will be much larger, the fibers will be positioned with robotic spines, the primary mirror will be segmented). The operations concept has therefore used information gathered from other astronomical facilities (e.g. Keck and HET for the segmented primary mirror).

We also introduced new pieces of software that will be developed to deal with the large number of targets expected in the survey, the large number of fibers in the focal surface, and the need for automatic operations. Prototypes for the Exposure Time Calculator and the Target Allocation Simulator are already in development to support the science team in the definition of the Design Reference Survey, which will describe the first years of operations of MSE with specific science goals in mind. More advanced software tools will need to be developed at a later stage by the partners, with support from the MSE Project Office, that has provided a framework of requirements to design these tools: the Observing Matrix Generator, responsible for the schedule, and the Automatic Data Reduction Pipeline, responsible for reducing and evaluating the data.

\acknowledgements{The Maunakea Spectroscopic Explorer (MSE) conceptual design phase was conducted by the MSE Project Office, which is hosted by the Canada-France-Hawaii Telescope (CFHT). MSE partner organizations in Canada, France, Hawaii, Australia, China, India, and Spain all contributed to the conceptual design. The authors and the MSE collaboration recognize the cultural importance of the summit of Maunakea to a broad cross section of the Native Hawaiian community.}

\bibliography{report} 
\bibliographystyle{spiebib} 

\end{document}